\documentclass[twocolumn,prl,amssymb,showpacs]{revtex4-1}
\usepackage{amsmath}
\usepackage{graphicx}
\usepackage{subfigure}

\def\vs{\medskip}

\begin{document}

\title{Nonlocality and entanglement in the XY model}

\author{J. Batle and M. Casas}
\email{E-mail address (JB): vdfsjbv4@uib.es}
\affiliation{Departament de F\'{\i}sica and IFISC-CSIC, Universitat de les Illes Balears, 07122 Palma de Mallorca, Spain}

\date{\today}

\begin{abstract}

Nonlocality and quantum entanglement constitute two special features of quantum systems of paramount
importance in quantum information theory (QIT). Essentially regarded as identical or equivalent for many years, 
they constitute different concepts.
Describing nonlocality by means of the maximal violation of two Bell inequalities, we study both entanglement and nonlocality 
for two and three spins in the XY model. Our results shed a new light into the description of nonlocality and the
possible information-theoretic task limitations of entanglement in an infinite quantum system.

\end{abstract}
\pacs{03.65.Ud; 03.67.Bg; 03.67.Mn; 64.70.Tg; 05.30.Rt}
\maketitle

Schr\"odinger 's reply \cite{Schro} to the paradox posed by Einstein, Podolsky and Rosen (EPR) \cite{EPR} 
motivated the modern notion of entanglement in a quantum system.
EPR suggested a description of nature, called ``local realism'', which assigned independent 
properties to distant parties of a composite physical system, 
to conclude that QM was an incomplete theory. Schr\"odinger, instead, did not recognise such conflict 
and regarded entanglement as {\it the} characteristic feature of QM.

The most significant progress toward the resolution of the EPR debate was made by Bell \cite{Bell}.
Bell showed that local realism, in the form of local variable models (LVM), implied
constraints on the predictions of spin correlations, known
as Bell inequalities. Spatially separated observers sharing an entangled state and performing
measurements on them may induce (nonlocal) correlations which cannot be simulated by local means
(violate Bell inequalities). This limitation to our physical understanding is nowadays exploited for 
implementing information-theoretic tasks.

Ever since Bell's contribution, entanglement and nonlocality were essentially identified
as the same thing. With the advent QIT, interest in
entanglement dramatically increased over the years for it lies at the basis of several 
important processes and applications which possess no classical counterpart \cite{Nielsen,Ekert}. 

Confusion between nonlocality and entanglement arose when the usefulness of
quantum correlations was put in doubt (see \cite{Gisin}).
The nonlocal character of entangled states was clear for pure states since all entangled pure
states of two qubits violate the CHSH inequality and are therefore nonlocal ({\it Gisin's theorem})
\cite{GisinTheorem}. However, the situation became more involved when Werner \cite{Werner} discovered that
while entanglement is necessary for a state to be nonlocal, for mixed states is not sufficient.

Entanglement is commonly viewed as a useful resource
for various information-processing tasks. Yet, there exist certain procedures, such as device-independent
quantum key distribution \cite{device} and quantum communication complexity problems \cite{comcomplex},
which can only be carried out provided the corresponding entangled states exhibit nonlocal correlations.
Therefore we are naturally led to the question whether nonlocality and entanglement constitute two different
resources.

The purpose of the present work is to shed some light upon the relation between entanglement
and nonlocality, through the maximal violation of a Bell inequality, in an infinite system, namely,
the XY model \cite{LSM} . 

The model Hamiltonian of the anisotropic one-dimensional
spin-$\frac{1}{2}$ XY model in a transverse magnetic field $h$ for $N$ particles is given by
\begin{equation} \label{equacioXY}
H=\sum_{i=1}^N [ (1+\gamma)S_x^{j}S_x^{j+1} + (1-\gamma)S_y^{j}S_y^{j+1} ]-h\sum_{i=1}^N S_x^{j},
\end{equation}
\noindent where $\sigma^j_u=2S_u^{j}$ ($u=x,y,z$) are the Pauli
spin-$\frac{1}{2}$ operators on site $j$, $\gamma \in [0,1]$ and $\sigma^{j+N}_u=\sigma^j_u$. 
The XY model (\ref{equacioXY}) for $N=\infty$ is completely
solved by applying a Jordan-Wigner transformation \cite{LSM,Barouch}, which maps the Pauli (spin 1/2) algebra into 
canonical (spinless) fermions. This model undergoes a paramagnetic-to-ferromagnetic
quantum phase transition (QPT) \cite{Sachdev} driven by the parameter $h$ at $h_c=1$ and $T=0$.

We will provide evidence for an anomaly that regards entanglement and nonlocality in the
XY model. To such end, we will consider the correlations existing between two sites or qubits (bipartite case)
and three sites or qubits (tripartite case).

\vs{\em Two qubits} The general two-site density matrix is expressed as
\begin{equation} \label{rho2}
\rho_{ij}^{(R)} = \frac{1}{4} \,
\Bigg[ \mathbb{I} + \sum_{u,v} T_{uv}^{(R)}
\sigma^i_u \otimes \sigma^j_v  \Bigg].
\end{equation}
\noindent $R=j-i$ indicates the distance between spins, $\{u,v\}$ denote any index of 
$\{\sigma_0,\sigma_x,\sigma_y,\sigma_z\}$, and $T_{uv}^{(R)} \equiv \langle \sigma^i_u \otimes \sigma^j_v \rangle$. 
Due to symmetry considerations, only
$\{T_{xx}^{(R)},T_{yy}^{(R)},T_{zz}^{(R)},T_{xy}^{(R)}\}$ do not vanish. Barouch {\it et al} \cite{Barouch} provided
exact expressions for two-point correlations, together with all the dynamics associated with an 
external $h(t)$. Let us consider the case where $h$ jumps from and initial value $h_0$ to a final
value $h_f$ at $t=0$ (the equilibrium case is easily recovered when $h_f=h_0$) and the $R=1$ configuration.
Following \cite{Barouch}, one obtains that
$T_{xx}^{(1)}=G_{-1},T_{yy}^{(1)}=G_{1},T_{zz}^{(1)}=G_0^2-G_1G_{-1}-S_1S_{-1}$ and $T_{xy}^{(1)}=S_1$, where
\begin{eqnarray}
G_R&=& \frac{\gamma}{\pi} \int_0^{\pi} d\phi \sin (R\phi)
\frac{\tanh \big[ \frac{1}{2}\beta \Lambda(h_0) \big]}{\Lambda(h_0)\Lambda^2(h_f)}\cr
&&\times [ \gamma^2\sin ^2\phi + (h_0-\cos \phi)(h_f - \cos \phi) \cr
&&-(h_0-h_f)(h_f-\cos \phi) \cos (2\Lambda(h_f)t) ] \cr
&&-\frac{1}{\pi} \int_0^{\pi} d\phi \cos (R\phi)
\frac{\tanh \big[ \frac{1}{2}\beta \Lambda(h_0) \big]}{\Lambda(h_0)\Lambda^2(h_f)}\cr
&&\times \big[ \{ \gamma^2\sin ^2\phi + (h_0-\cos \phi)(h_f - \cos \phi)\}(\cos \phi - \cr
&& h_f)-(h_0-h_f)\gamma^2\sin ^2\phi \cos (2\Lambda(h_f)t) ],\\ \label{G}
&& \cr
S_R&=&\frac{\gamma (h_0-h_f)}{\pi}\int_0^{\pi} d\phi\sin (R\phi) \sin \phi
\frac{\sin [ 2\Lambda(h_f)t ]}{\Lambda(h_0)\Lambda(h_f)}, \label{G}
\end{eqnarray}
\noindent with $\Lambda(h)=[\gamma^2\sin^2\phi + (h-\cos \phi)^2]^{1/2}$. $G_R$ is the two-point correlator
appearing in the Wick theorem calculations, and $M_z=\frac{1}{2}G_0$.

Most of our knowledge on Bell inequalities and their quantum mechanical
violation is based on the CHSH inequality \cite{GisinTheorem}. With two dichotomic observables per party, it is the
simplest nontrivial Bell inequality for the bipartite case with
binary inputs and outcomes. Quantum mechanically, these observables reduce
to ${\bf A_j}({\bf B_j})=\bf{a_j}(\bf{b_j}) \cdot \bf{\sigma}$, where $\bf{a_j}(\bf{b_j})$
are unit vectors in $\mathbb{R}^3$ and $\bf{\sigma}=(\sigma_x,\sigma_y,\sigma_z)$ the Pauli matrices.
Violation of CHSH inequality requires $Tr(\rho_{ij}^{(R)} B_{CHSH})$, that is, the expectation value of
the operator $B_{CHSH}$
\begin{equation}
{\bf A_1}\otimes {\bf B_1} + {\bf A_1}\otimes {\bf B_2} 
+ {\bf A_2}\otimes {\bf B_1}  -  {\bf A_2}\otimes {\bf B_2} 
\end{equation} 
\noindent to be greater than 2. We shall take the optimum value over all $\{ \bf{a_j},\bf{b_j} \}$ as a proper
measure for nonlocality in our state $\rho_{ij}^{(R)}$ (\ref{rho2}).
This procedure is presented in detail elsewhere \cite{futurjo}.
Given a general two qubit state $\rho$ in the usual computational basis, we change it into
the well known Bell basis
$\{ |\Phi^{+}\rangle,|\Phi^{-}\rangle,|\Psi^{+}\rangle,|\Psi^{-}\rangle \}$. The ensuing matrix
$\rho = \rho_{\parallel} + \rho_{\perp}$ is decomposed into two contributions, where only
terms in $\rho_{\parallel}$
\begin{equation} \label{rhoBell}
 \left( \begin{array}{cccc}
\rho_{11} & i\rho^{I}_{12} & i\rho^{I}_{13} & \rho^{R}_{14}\\
-i\rho^{I}_{12} & \rho_{22} & \rho^{R}_{23} & i\rho^{I}_{24}\\
-i\rho^{I}_{13} & \rho^{R}_{23} & \rho_{33} & i\rho^{I}_{34}\\
\rho^{R}_{14} & -i\rho^{I}_{24} & -i\rho^{I}_{34} & \rho_{44} \end{array} \right)
\end{equation}
\noindent contribute to $Tr(\rho B_{CHSH})$. In the XY model,
state (\ref{rho2}) is almost Bell-diagonal except for $\rho^{I}_{12}=\frac{1}{2}T_{xy}^{(R)}$, which is null
in equilibrium ($h_f=h_0$).

\begin{figure}[htbp]
\begin{center}
\includegraphics[width=8.6cm]{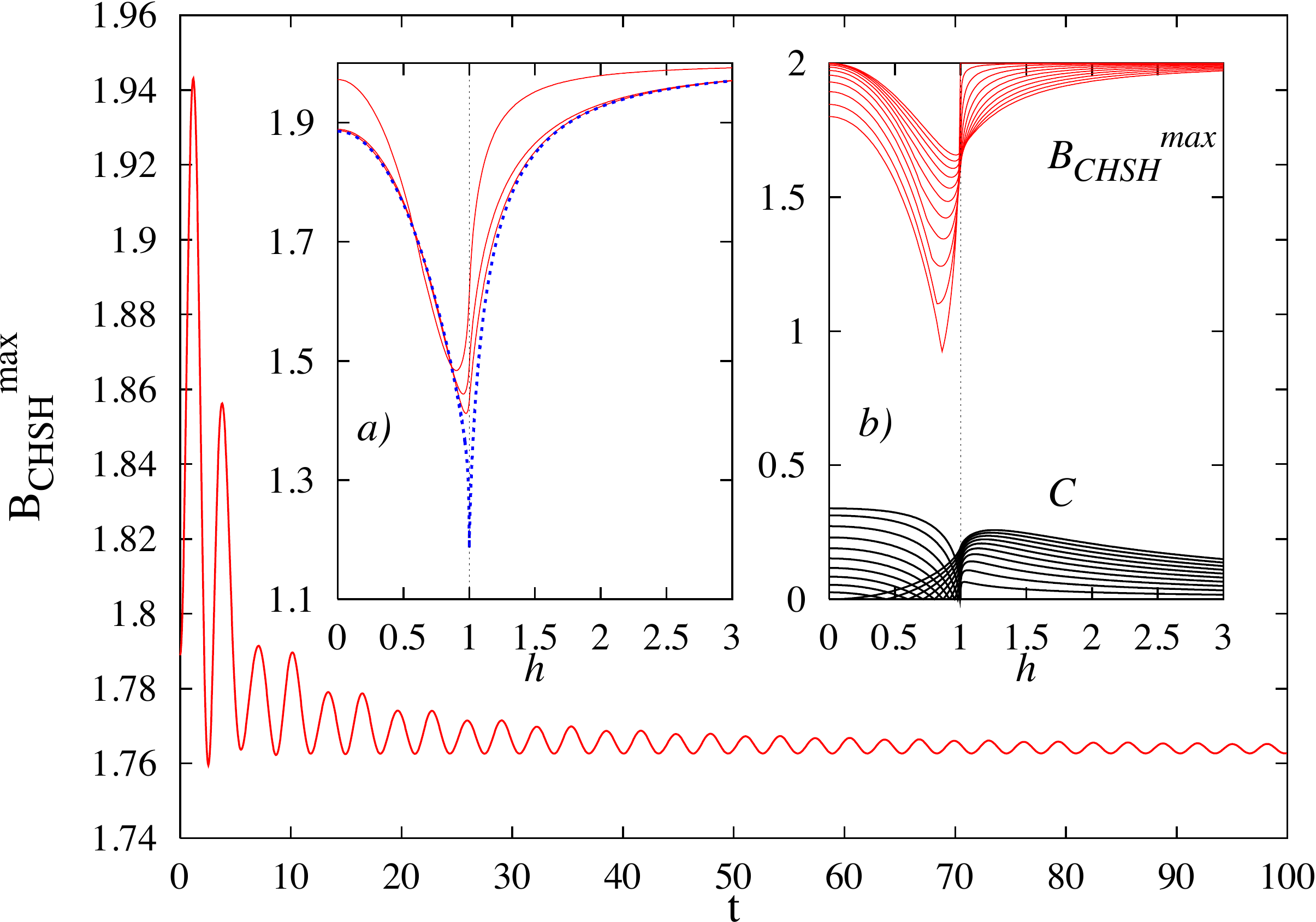}
\caption{(colour online). The oscillating curve depicts $B_{CHSM}^{\max}$ (\ref{CHSHmax}) versus time (in seconds)  
for $h_0=0.5,h_f=0$ and $\gamma=0.5$ for the evolved state $\rho_{ij}^{(R)}(t)$ (\ref{rho2}) at T=0. Final and equilibrium 
values do not coincide ({\it non-ergodic}). Inset a) depicts $B_{CHSM}^{\max}$ versus $h$ for $\gamma=0.5$ and 
different distances between spins $R=1,2,3$ and $\infty$ (dotted line). Though no Bell violation is observed, $B_{CHSM}^{\max}$ exhibits a long range order. 
Inset b) depicts equilibrium nonlocality (upper curves) and concurrence (lower curves) measures 
versus $h$ for $\gamma \in [0,1]$. The isotropic case $\gamma=0 $ collapses to 2 for $h \geq 1$ . See text for details.}
\label{fig1}
\end{center}
\end{figure}

\noindent Optimization is carried out and after some algebra, we obtain
$2\sqrt{2}\sqrt{(\rho_{11}-\rho_{44})^2+(\rho_{22}-\rho_{33})^2+4(\rho^{I}_{12})^2}$, with the
diagonal elements of (\ref{rhoBell}) arranged so that $\rho_{11}>\rho_{22}>\rho_{33}>\rho_{44}$. The
specific form for our state (\ref{rho2}) reads as
\begin{eqnarray} \label{CHSHmax}
 B_{CHSM}^{\max} &\equiv& \max_{\bf{a_j},\bf{b_j}}\,\,Tr (\rho_{ij}^{(R)} B_{CHSH}) \cr
 &=& 2 \sqrt{ \Vert {\bf T^{(R)}} \Vert^2 -
\, \min \,_{xyz}
 + 2 \big[ T_{xy}^{(R)} \big]^2},
\end{eqnarray}
\noindent where ${\bf T^{(R)}}=(T_{xx}^{(R)},T_{yy}^{(R)},T_{zz}^{(R)})$ and
$\min_{xyz} \equiv \min \big( \big[ T_{xx}^{(R)} \big]^2,
\big[ T_{yy}^{(R)} \big]^2, \big[ T_{zz}^{(R)} \big]^2 \big)$. Fig. 1 depicts the evolution of nonlocality
(\ref{CHSHmax}) for the case $\gamma=0.5$ and ($h_0=0.5,h_f=0$). For $\gamma=1$ (Ising) and ($h_0=h_c,h_f=0$) (\ref{CHSHmax}) 
would oscillate indefinitely. This nonlocality measure oscillates around a value distinct from the one expected. Just as in the case of the
$M_z$ in \cite{Barouch}, or entanglement in the XY model \cite{Aditi} (as measured by the concurrence $C$),
(\ref{CHSHmax}) does not reach its equilibrium value, which entails that nonlocality is also a non-ergodic quantity.

The equilibrium case ($h_f=h_0$) is considered in Fig. 1a. $B_{CHSM}^{\max}$ evolves for different configurations up to $R=\infty$ \cite{Barouch}. 
It is apparent that nonlocality exhibits a long range behavior, while
$C$ rapidly tends to zero \cite{osterlochnature}.
Comparison with entanglement appears in Fig. 1b for all
$\gamma$-anisotropies. The usefulness of (bi)entanglement between spins in the XY model is questioned in quantum information processing
by the fact that the concomitant correlations never violate a Bell inequality ($B_{CHSM}^{\max} \le 2\,\, \forall \,h,\gamma$).

All previous quantities are ultimately described in terms of several $G_R$, so that they all diverge
at $h=1$ in the same manner. Let us consider for simplicity
$M_z(h)=\frac{1}{2}G_0=\frac{\partial}{\partial h} \frac{1}{2\pi}\int_0^{\pi} d\phi [\gamma^2\sin^2\phi + (h-\cos \phi)^2]^{1/2}$.
For $\gamma=1$ we have $M_z(h)=\frac{\partial}{\partial h} \big( \frac{2(h+1)}{2\pi} E \big[\frac{2\sqrt{h}}{h+1} \big] \big)=
\frac{1}{2\pi} \big[ \frac{h-1}{h} K \big(\frac{2\sqrt{h}}{h+1}\big) + \frac{h+1}{h} E \big(\frac{2\sqrt{h}}{h+1}\big) \big]$,
where $K(E)$ is the complete elliptic integral of the first(second) kind. $\frac{d}{dh}M_z$ diverges logarithmically at $h=1$ following
the divergence of $K$, and so does $C$ and $B_{CHSM}^{max}$, including non-equilibrium ($t=\infty$) values.
Therefore entanglement and nonlocality exhibit one additional feature beside
non-ergodicity: they both signal the presence of a QPT at T=0.

\vs{\em Three qubits}  Nonlocality in the three qubit case is explored through the violation of the Mermin inequality \cite{Mermin}. 
The Mermin inequality reads as
$Tr(\rho B_{Mermin}) \leq 2$, where $B_{Mermin}$ is the Mermin operator
\begin{equation} \label{Mermin}
 B_{Mermin}=B_{a_{1}a_{2}a_{3}} - B_{a_{1}b_{2}b_{3}} - B_{b_{1}a_{2}b_{3}} - B_{b_{1}b_{2}a_{3}},
\end{equation}
\noindent with $B_{uvw} \equiv \bf{u} \cdot \bf{\sigma} \otimes \bf{v} \cdot \bf{\sigma} \otimes \bf{w} \cdot \bf{\sigma}$ 
with ${\bf \sigma}=(\sigma_x,\sigma_y,\sigma_z)$ being the usual Pauli matrices, and $\bf{a_j}$ and $\bf{b_j}$ unit vectors
in $\mathbb{R}^3$. Notice that GHZ states maximally violate the Mermin inequality. 
As usual, we will employ 
\begin{equation} \label{MerminMax}
 Mermin^{\max} \equiv \max_{\bf{a_j},\bf{b_j}}\,\,Tr (\rho B_{Mermin})
\end{equation}
\noindent as a measure for the nonlocality of the state $\rho$.

The XY model is completely solvable, a fact that allows us to compute -- as in the previous case of two sites -- the reduced density matrix
for three spins without the explicit construction of the global infinite state of the system. The reduced state of three spins reads as
\begin{equation} \label{rho3}
\rho_{ijk}^{(a,b)} = \frac{1}{8} \,
\Bigg[ \mathbb{I} + \sum_{u,v,w} T_{uvw}^{(a,b)}
\sigma^i_u \otimes \sigma^j_v \otimes \sigma^k_w \Bigg],
\end{equation}
\noindent where $i<j<k$ indicate the positions of the three spins and $a=j-i, b=k-j$ their relative distances.
$\{u,v,w\}$ denote indexes of the Pauli matrices $\{\sigma_0,\sigma_x,\sigma_y,\sigma_z\}$,
and $T_{uvw}^{(a,b)} \equiv \langle \sigma^i_u \otimes \sigma^j_v \otimes \sigma^k_w \rangle_{ab}$.
Similarly to the calculation of the two-spin correlations computed by Barouch {\it et al} \cite{Barouch},
based in turn on the work by Lieb {\it et al} \cite{LSM},
we extend them to the three party case by using the well known Wick theorem in quantum field theory.
Due to the symmetry of the XY model, some of them vanish. Furthermore, as far as nonlocality is concerned,
among those correlations who survive only four of them contribute to (\ref{MerminMax}), namely,
$T_{xxz}^{(a,b)},T_{xzx}^{(a,b)},T_{zxx}^{(a,b)}$ and $T_{zzz}^{(a,b)}$.
These three-spin correlation functions $T_{xzx}^{(a,b)}, T_{xxz}^{(a,b)}$ and $T_{zzz}^{(a,b)} $ are given by
\begin{equation}\label{det4}
\left| \begin{array}{cccccc}
G_{-1} & \dots & G_{-a+1} & G_{-a-1} & \dots & G_{-a-b} \\
\vdots &       & \vdots   & \vdots   &       & \vdots \\
G_{a-2}& \dots & G_{0}   & G_{-2}    & \dots & G_{-b-1} \\
G_{a} & \dots & G_{2} & G_{0}  & \dots & G_{-b+1} \\
\vdots &       & \vdots   & \vdots   &       & \vdots \\
G_{a+b-2} & \dots & G_{b} & G_{b-2}  & \dots & G_{-1} \end{array} \right|,
\end{equation}
\begin{figure}[htbp]
\begin{center}
\includegraphics[width=8.6cm]{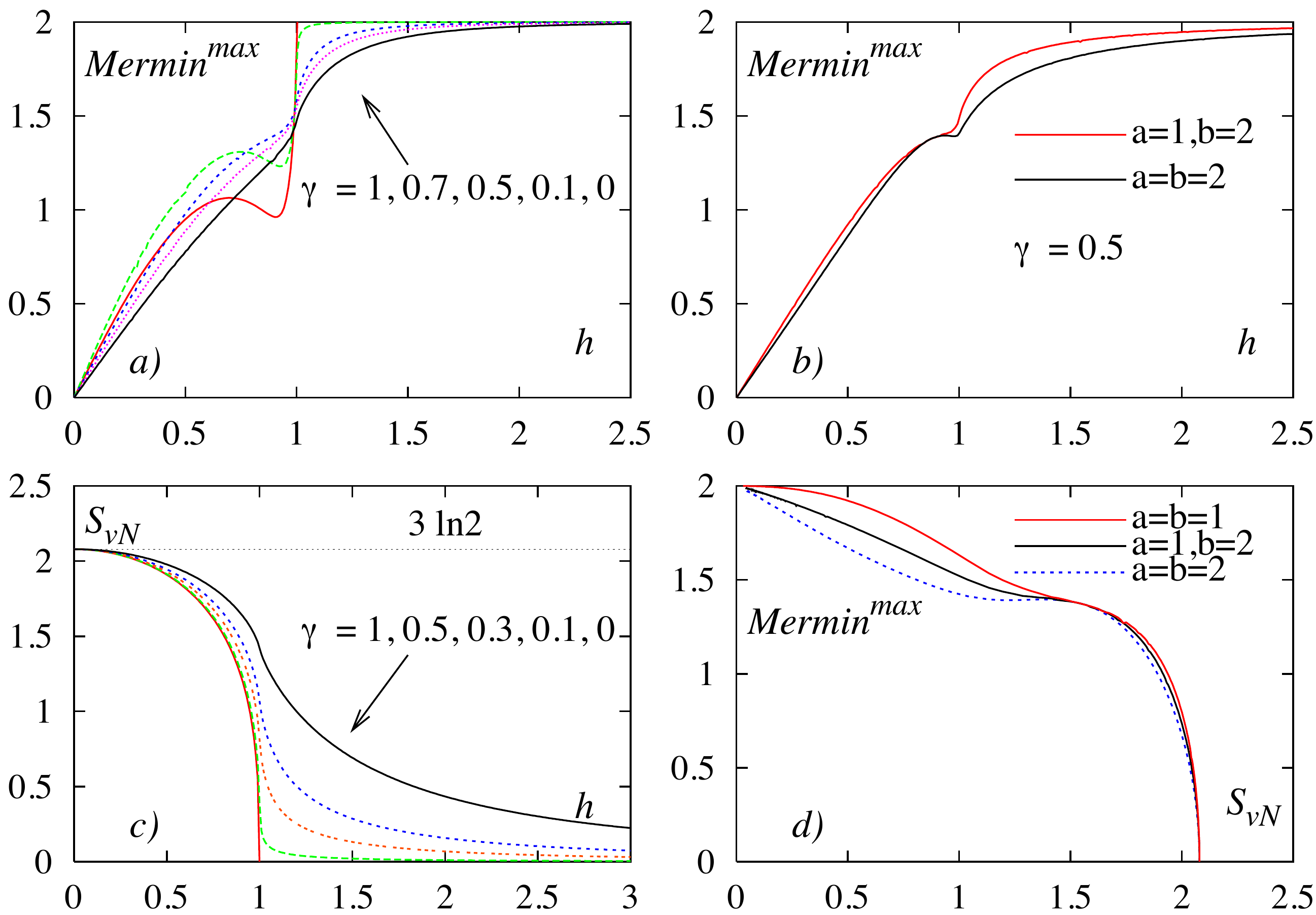}
\caption{(colour online). Fig. a) depicts the value of $Mermin^{\max}$ versus $h$ (\ref{MerminMax}) for several $\gamma$ 
values for the $a=b=1$ block configuration. Fig. b) shows similar curves for $a=1,b=2$ and $a=2,b=2$, with $\gamma=0.5$. Fig. c) 
depicts the evolution of multipartite entanglement $S_{vN}$ ($a=b=1$) versus $h$ for several values of the anisotropy. Fig. d) plots 
$Mermin^{\max}$ versus $S_{vN}$ for $\gamma=0.5$ and all previous configurations $a=b=1$, $a=1,b=2$ and $a=b=2$. Notice the apparent monotonic 
decreasing evolution. See text for details.}
\label{fig2}
\end{center}
\end{figure}
\begin{equation}\label{det6}
\left| \begin{array}{cccc}
G_{-1} & \dots & G_{-a}   & G_{-a-b}   \\
G_{0}   & \dots & G_{-a+1} & G_{-a-b+1} \\
\vdots  & \ddots & \vdots  & \vdots \\
G_{a-2} & \dots & G_{-1} & G_{-b-1} \\
G_{a+b-1} & \dots & G_{b} & G_{0} \end{array} \right|,
\left| \begin{array}{ccc}
G_{0} & G_{-a} & G_{-a-b}  \\
G_{a}  & G_{0} & G_{-b}  \\
G_{a+b}  & G_{b} & G_{0} \end{array} \right|
\end{equation}
\noindent respectively, with $T_{zxx}^{(a,b)}=T_{xxz}^{(b,a)}$ due to translational symmetry. 

Correlations between parties strongly depend on their relative positions $a$ and $b$.
We shall distinguish two types: i) the one forming a block of three in consecutive sites ($a=b=1$), as well as a centered
configuration ($a=b=2$), and ii) two nearest-neighbors spins plus an additional one ($a=1,b=2$).

Optimization of $Mermin^{\max}$ (\ref{MerminMax}) for any configuration of the spins along the chain is carried out as follows.
Once the observers' settings $\{\bf{a_j},\bf{b_j}\}$ are parameterized in spherical coordinates $(\sin\theta_k \cos\phi_k,\sin\theta_k \sin\phi_k,\cos\theta_k)$,
the problem consists in finding the supremum of (\ref{MerminMax}) over the set of $\{k=1..12\}$ possible angles. 
Here, too, we will consider ground state nonlocality.

Let us consider the block configuration $a=b=1$. The orientation of the settings $\{{\bf a_j},{\bf b_j} \}$ that
maximizes $Mermin^{\max}$ (\ref{MerminMax}) is such that ($x-z$ plane only, that is, we deal we {\it real qubits} for $u_j^y=0 \, \forall {\bf u}$)
$\{{\bf a_3}=-{\bf a_1},{\bf b_1}={\bf a_2},{\bf b_2}=-{\bf a_1},{\bf b_3}=-{\bf a_2} \}$ for the whole
range of $h$. Explicitly, $Mermin^{\max} = max_{\theta_{a_1},\theta_{a_2}}  \big[ a_2^z\big((a_2^z)^2-3(a_1^z)^2
\big)T_{zzz}^{(1,1)}+\big( a_2^z\big((a_2^x)^2-(a_1^x)^2\big)-2a_1^z a_1^x a_2^x \big) \big( T_{zxx}^{(1,1)}
+ T_{xzx}^{(1,1)} + T_{xxz}^{(1,1)} \big) \big]$. The ensuing analytic form of $Mermin^{\max}$ versus $h$
-- a complicated rational function with radicals-- is calculated by recourse to convex optimization. 
For the sake of generality, let us also consider the centered configuration $a=b=2$. After some algebra, we obtain
that one disposition of the observers that provides
an analytical expression for a lower bound to (\ref{MerminMax}) is given by (again in the $x-z$ plane)
$\{ {\bf a_1}=(\sin\theta_{a_1},\cos\theta_{a_1}),{\bf a_2}=(0,-1),{\bf a_3}=(\sin\theta_{a_1},-\cos\theta_{a_1}),
{\bf b_1}={\bf a_3},{\bf b_2}=(1,0),{\bf b_3}=-{\bf a_1} \}$.
Hence, we obtain that
$max_{\theta_{a_1}} \big[ 2\cos^2\theta_{a_1}T_{zzz}^{(2,2)} +
2\sin\theta_{a_1}\cos\theta_{a_1}T_{zxx}^{(2,2)} -2\sin^2\theta_{a_1}T_{xzx}^{(2,2)} + 2\sin\theta_{a_1}\cos\theta_{a_1}T_{xxz}^{(2,2)} \big]
\leq Mermin^{\max} \leq \sqrt{4\big(T_{zzz}^{(2,2)}\big)^2 + 4\big(T_{zxx}^{(2,2)}\big)^2 + 4\big(T_{xzx}^{(2,2)}\big)^2 + 4\big(T_{xxz}^{(2,2)}\big)^2 }$.
The analytic form of the lower bound -- not provided here -- is of the same nature of that of the $a=b=1$ case. In point of fact, the lower bound
becomes an equality for all $(a,b)$ shortly before $h=1$. Additionally, the upper bound also applies to all configurations.

Fig. 2a,2b present the situation, where $Mermin^{\max}$ is depicted for different
values of the $\gamma$-anisotropy and different configurations of the spins. Numerical calculations perfectly agree with 
the corresponding analytic expressions.
As $h$ grows, the state $\rho_{ijk}^{(a,b)}$ approaches $|\downarrow \downarrow \downarrow \,\,\rangle \langle \, \, \downarrow \downarrow \downarrow|$
 as expected (ferromagnetic phase), and {\it never violates the Mermin inequality}, 
 which entails an inherent limitation to the usefulness of entanglement itself.

Characterization of entanglement is of paramount relevance
in QIT \cite{RMPamico}, yet no operational measure is available to date that
quantifies genuine multipartite entanglement. 
We will nevertheless employ the sum of the von Neumann
entropy of the reduced density matrices of the three spins of $\rho_{ijk}^{(a,b)}$, $S_{vN}=3 S_{vN}(\rho_i), $ with $\rho_{i}= \frac{1}{2}
(\mathbb{I} + \langle \sigma_z \rangle  \sigma^i_z )$, $\langle \sigma_z \rangle = G_0 = 2M_z$. 
We shall consider ground state entanglement (T=0), as well. The specific form of $S_{vN}$ is depicted in Fig. 2c for several
values of the $\gamma$-anisotropy as a function of $h$. The monotonic decreasing tendency of entanglement
is apparent for any $\gamma$-value. As $h$ grows, the fidelity between $\rho_{ijk}^{(a,b)}$ steadily tends to 1.

Remarkably, we observe opposite tendencies revealed by entanglement $S_{vN}$ and nonlocality (\ref{MerminMax}). Recall
that our model undergoes a second order QPT in the ground energy. In point of fact, since 
$\frac{d}{dh}S_{vN}   \propto   \frac{d}{dh}M_z$, this measure of multipartite entanglement (its derivative) diverges logarithmically 
at $h_c=1$. Surprisingly, our measure of nonlocality (\ref{MerminMax}) also displays such
divergence, along with its bounds. In Fig. 2d we show the dependency of $Mermin^{max}$ versus entanglement $S_{vN}$ for 
several values of $\gamma$ in the ($a=1,b=1$) configuration. Finally, we encounter for the first time a multipartite system where 
not only is entanglement (its first derivative) a good indicator of a QPT, but also nonlocality. 

\vs{\em Conclusions}  We have studied how nonlocality --measured by the maximal violation of a Bell inequality-- compares to entanglement in a 
condensed matter system. Although two instances (two and three sites) have been considered, our results may properly generalize 
to any block of spins. 
For the bipartite case, we have computed the exact value of $B_{CHSM}^{\max}$ during time evolution and in equilibrium. In either cases 
our nonlocality measure (\ref{CHSHmax}) displays a non-ergodic behavior, is able to detect a QPT, and limits the QIT-related tasks 
involving bipartite entanglement along the infinite chain since no Bell inequality is violated. 
A similar situation occurs in the tripartite case, where non-violation of local realism in the XY model takes place as well. Also, 
entanglement and nonlocality both indicate a QPT yet exhibit opposite evolutions in the phase diagram. 
Finally, nonlocality can also constitute a complementary resource in infinite quantum systems.

\vs{\em Acknowledgements } J. Batle acknowledges fruitful discussions with J. Rossell\'{o}. M. Casas acknowledges 
partial support by the MEC grant FIS2005-02796 (Spain) and FEDER (EU).

\end{document}